\documentclass[9pt,onecolumn,twoside]{pnas-new}

\templatetype{pnasmathematics} 

\title{Atmospheric stability sets maximum moist heat and convection in the midlatitudes}

\author[a,1]{Funing Li}
\author[a]{Talia Tamarin-Brodsky}  

\affil[a]{Department of Earth, Atmospheric, and Planetary Sciences, Massachusetts Institute of Technology, Cambridge, MA 02139, USA} 

\leadauthor{Li} 

\authorcontributions{F.L. and T.T-B. designed research and wrote the paper; F.L. performed research.}
\authordeclaration{The authors declare no competing interest.} 
\correspondingauthor{\textsuperscript{1}To whom correspondence should be addressed. E-mail: lifuning1991@gmail.com}


\begin{abstract}
Extreme near-surface moist heat and severe convective storms are among the leading causes of weather-related damages worldwide. Here, we show that episodes of extreme moist heat and severe convection frequently co-occur across midlatitude land regions, and develop a theoretical framework that links their maximum potential intensities to preexisting low-level energy inversions. By accounting for the stored-energy nature of midlatitude severe convection, where moist heat and atmospheric instability accumulate before convection initiates, our work advances the understanding of convective constraints on extreme heat events. The theory identifies low-level inversions as a critical factor shaping compound extreme heat and convective weather risks, and offers a pathway for improving the modeling and future projection of these events. 
\end{abstract}

\dates{This manuscript was compiled on \today}

\begin{document}

\maketitle
\thispagestyle{firststyle}
\ifthenelse{\boolean{shortarticle}}{\ifthenelse{\boolean{singlecolumn}}{\abscontentformatted}{\abscontent}}{}


Moist heatwaves, which combine the effects of high temperature and humidity, pose significant risks to public health and societal outcomes \cite{basu_etal_2002,davis_etal_2003,basu_2009,sherwood_huber_2010,davis_etal_2016,buzan_huber_2020,de_etal_2021,kong_huber_2022wbgt,lesk_etal_2022,saeed_etal_2022,baldwin_etal_2023,vanos_etal_2023,diaz_etal_2023,guo_etal_2024,matthews_etal_2025}. While previous research has primarily focused on the dynamics and changes of moist heat in tropical and subtropical regions \cite{mishra_etal_2020,byrne_2021,raymond_etal_2021,zhang_etal_2021_heat,duan_etal_2024,zhang_etal_2024,duan_etal_2024jcli}, the drivers and characteristics of extreme moist heat over midlatitude continents remain poorly understood. As climate change is driving severe moist heat risks poleward into the midlatitudes \cite{matthews_etal_2017,coffel_etal_2017,tuholske_etal_2021,raymond_etal_2020,rogers_etal_2021,romps_lu_2022,vecellio_etal_2023,zhang_etal_2023}, advancing our theoretical understanding of the limits of maximum moist heat in these regions has become increasingly urgent. Our goal in this study is to determine what sets the maximum moist heat over midlatitude land. 

\begin{figure*}[tbhp]
\centering
\centerline{\includegraphics[width=1.02\linewidth]{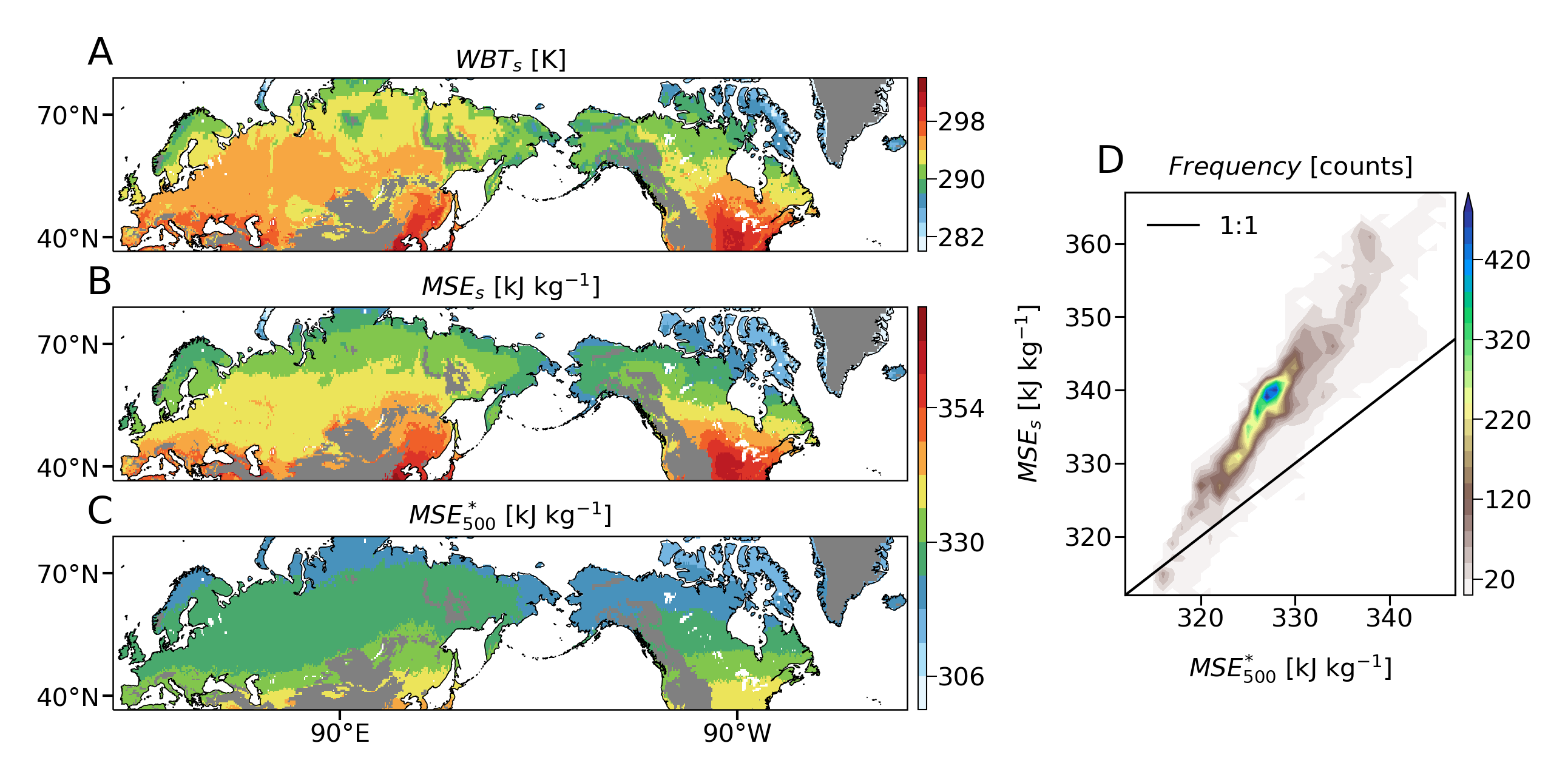}}
\caption{Extreme moist heat. ($A$): Annual maximum near-surface wet-bulb temperature ($WBT_s$). ($B$): Annual maximum near-surface moist static energy ($MSE_s$). ($C$): 500-hPa saturated moist static energy ($MSE_{500}^*$) associated with annual maximum $MSE_s$. ($D$): Joint histogram of annual maximum $MSE_s$ and the associated $MSE_{500}^*$, with a bin size of 1$\times$1 kJ kg$^{-1}$. Results are historical means based on ERA5 reanalysis data during 1980--2022 for land between 35$^{\circ}$N and 75$^{\circ}$N at elevations lower than 1000 m.}
\label{fig_01}
\end{figure*}

Near-surface wet bulb temperature (WBT$_s$) is a key measure of moist heat stress, as it combines the influence of temperature and humidity \cite{haldane_1905,sherwood_huber_2010,buzan_etal_2015,buzan_huber_2020,raymond_etal_2020,zhang_etal_2021_heat,vecellio_etal_2023,kong_huber_2023}. By definition, WBT represents the temperature that an air parcel would attain if cooled adiabatically to saturation at constant pressure by evaporating water into it \cite{haldane_1905,davies_2008,stull_2011}. During this adiabatic process, the air parcel conserves its moist static energy (MSE) or the equivalent potential temperature, and thus WBT$_s$ can be directly derived assuming the conservation of near-surface MSE (MSE$_s$) \cite{fischer_Knutti_2013,zhang_etal_2021_heat,raymond_etal_2021,kong_huber_2023,ivanovich_etal_2024}. Unlike the non-linear WBT formula, MSE is a linear function of temperature, specific humidity, and geopotential height, making it simple to use \cite{raymond_etal_2021,chavas_peters_2023}. Accordingly, this study employs MSE$_s$, defined as the sum of sensible heat, latent heat, and geopotential energy at 2 meters above the ground, to measure near-surface moist heat. Overall, MSE$_s$ reasonably represents moist heat and, similar to the annual maximum WBT$_s$ (Fig. \ref{fig_01}$A$), the annual maximum MSE$_s$ recovers well the spatial distribution of extreme moist heat over midlatitude land (Fig. \ref{fig_01}$B$) and aligns closely with WBT$_s$ (Fig. S1; detailed in $Materials$ $and$ $Methods$). In this work, we use the 3-hourly ERA5 reanalysis data \cite{NCAR_RDA_ERA5,Hersbach_etal_2020} during 1980--2022 for land grids between 35$^{\circ}$N and 75$^{\circ}$N with elevation lower than 1000 m (detailed in $Materials$ $and$ $Methods$). Our analysis focuses on the historical annual maximum MSE$_s$ to explore the theoretical constraint on the maximum moist heat over midlatitude land.   

Previous studies have commonly assumed a strong coupling between the near-surface atmosphere and the free troposphere during extreme heat through deep moist convection, such that MSE$_s$ is constrained by free troposheric saturated moist static energy (MSE$^*$) given the state of convective quasi-equilibrium \cite{emanuel_etal_1994,neelin_zeng_2000,williams_etal_2009,sherwood_huber_2010,byrne_OGorman_2013,singh_ogorman_2013,zamora_etal_2016,zhang_Stephan_2020,zhang_etal_2021_heat,byrne_2021,buzan_huber_2020}. Under the assumption of a moist-neutral atmospheric column, this framework has been widely used to study extreme heat in the tropics \cite{byrne_2021,zhang_etal_2021_heat,zhang_etal_2024} or the global mean state \cite{buzan_huber_2020}. With the neutrality assumption, a recent study \cite{zhang_boos_2023} has applied this framework to investigate extreme dry heat over midlatitude continents, suggesting that MSE$_s$ is limited by 500-hPa MSE$^*$ (i.e., MSE$_s\leq$MSE$^*_{500}$), thereby setting an upper bound for maximum surface air temperature. We first test the validity of this theory by comparing MSE$_s$ with MSE$^*_{500}$ during the annual maximum moist heat over midlatitude land. Our analysis reveals that while the theory (and the neutrality assumption) might be more correct in the dry limit \cite{zhang_boos_2023}, it does not hold for extreme moist heat, as MSE$_s$ is consistently and significantly greater than MSE$^*_{500}$ in nearly all cases (more than 96\% of the midlatitude land grid points; Fig. \ref{fig_01}$B$--$D$). Furthermore, the distribution of MSE$^*_{500}$ is largely zonally uniform (Fig. \ref{fig_01}$C$), which does not capture the strong zonal variation observed in the MSE$_s$ pattern (Fig. \ref{fig_01}$B$).

These results indicate a violation of the moist neutrality assumption because, unlike quasi-equilibrium convection in the tropics, severe convection over midlatitude continents often involves a strong accumulation of convective energy \cite{doswell_1987,Doswell_2001,chaboureau_etal_2004,bechtold_etal_2004,Zipser_etal_2006,Agard_Emanuel_2017,Emanuel_2023,lafleur_etal_2023,Tuckman_etal_2023,Tuckman_Emanuel_2024}, which may interact with surface heating in distinct ways. Although this stored-energy nature of midlatitude severe convection has been well recognized within the severe weather community \cite{Emanuel_1994,Markowski_Richardson_2011_book}, it has not been $quantitatively$ formulated to constrain either the maximum potential intensity of convection \cite{Agard_Emanuel_2017,Emanuel_2023,Tuckman_etal_2023,Tuckman_Emanuel_2024} or the maximum surface heat available at the onset of convection \cite{zhang_boos_2023}. Here, building on the stored-energy energy of severe convection and the common assumption that convective precipitation limits surface heat, we propose a new theoretical framework that accounts for the onset of convection, enables surface heat and convective instability to evolve beyond the neutral condition, and ultimately provides a tight constraint on both near-surface moist heat and potential convection over midlatitude land. 

In the main text, we present and validate our theoretical framework within current climate using ERA5 reanalysis data. We conclude by discussing the important implications of our results for predicting and understanding the maximum intensities of severe convection and near-surface moist heat in future climates.

\section*{Concurrent moist heat and convection extremes}

\begin{figure*}[tbhp]
\centering
\centerline{\includegraphics[width=1.02\linewidth]{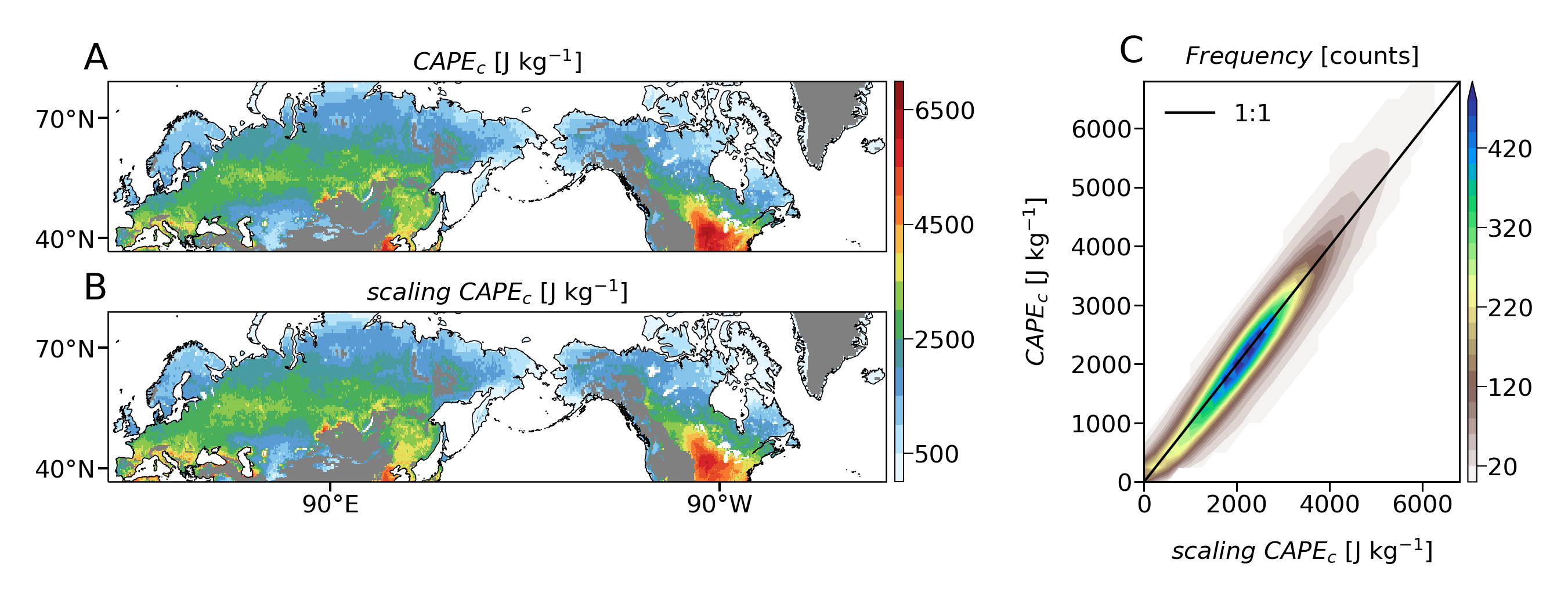}}
\caption{Extreme convective instability. ($A$): Critical CAPE ($CAPE_c$), defined as CAPE at the time of annual maximum $MSE_s$. ($B$): Theoretical scaling for $CAPE_c$ following Eq.\ref{scaling-cape}. ($C$): Joint histogram of $CAPE_c$ and scaling $CAPE_c$, with a bin size of 250$\times$250 J kg$^{-1}$. Results are historical means based on ERA5 reanalysis data during 1980--2022 for land between 35$^{\circ}$N and 75$^{\circ}$N at elevations lower than 1000 m.}
\label{fig_02}
\end{figure*}

We first demonstrate that deep convection (Fig. \ref{fig_02}) occurs concurrently with the annual maximum moist heat (Fig. \ref{fig_01}), supporting the hypothesis that convective instability acts to terminate heatwaves and constrain maximum near-surface heat \cite{buzan_huber_2020,zhang_boos_2023,sauter_etal_2023a,sauter_etal_2023b,duan_etal_2024}. The potential intensity of convection is measured by convective available potential energy (CAPE; detailed in $Materials$ $and$ $Methods$), which is defined as the vertical integral of positive parcel buoyancy through the deep troposphere, with values above 1000 J kg$^{-1}$ typically deemed sufficient to support strong to severe convection \cite{nwc-cape}. We define CAPE at the time of the annual maximum MSE$_s$ as critical CAPE (denoted CAPE$_c$), which measures the potential convection that terminates the annual maximum moist heat. The CAPE$_c$ over most midlatitude land regions exceeds 1000 J kg$^{-1}$, with values ranging from 3000 to 4000 J kg$^{-1}$ across western Europe extending to Northeast China, and reaching 6000 J kg$^{-1}$ or higher over eastern China and central North America (Fig. \ref{fig_02}$A$). The high values of CAPE$_c$ represent substantial convective instability in the atmosphere during extreme moist heat, indicating a significant departure from the neutrality assumption. Meanwhile, critical convective inhibition (denoted CIN$_c$; detailed in $Materials$ $and$ $Methods$), a measure of the vertically integrated energy barrier in the lower free troposphere that suppresses the release of CAPE, is sufficiently low (generally less than 25 J kg$^{-1}$) to permit the initiation of free convection (Fig. S2$A$). 

These high-CAPE, low-CIN environments largely ensure the occurrence of deep convection that terminates extreme near-surface heat, as further evidenced by substantial precipitation following the hottest MSE$_s$ day across most midlatitudes regions (Fig. S2$E$). In fact, the deep convection that terminates the maximum moist heat is likely among the most intense convection of the year across many midlatitude land areas, as CAPE$_c$ (Fig. \ref{fig_02}$A$) aligns closely with the annual maximum CAPE (Fig. S2$B$) and the precipitation following peak MSE$_s$ generally ranks in the high percentiles (top 10--20\%) of daily precipitation during the year (Fig. S2$F$). This alignment is particularly evident in Europe and central North America, where the difference between CAPE$_c$ and the annual maximum CAPE is less than 250 J kg$^{-1}$ (Fig. S2$C$). In these regions, over 50\% of the years during 1980--2022 feature simultaneous annual maxima of CAPE and MSE$_s$ (Fig. S2$D$), emphasizing the frequent co-occurrence of extreme moist heat and intense convection in the midlatitudes. 

Note that several regions including the land east of the Caspian Sea, south of the Mediterranean Sea, and along the west coast of the United States, exhibit high MSE$_s$ maxima (Fig. \ref{fig_01}$A$--$B$) but only modest CAPE$_c$ (Fig. \ref{fig_02}$A$), however, the relatively high CIN$_c$ (up to 100 J kg$^{-1}$ or greater) found in these regions (Fig. S2$A$) likely inhibits convective initiation and strongly limits precipitation there (Fig. S2$E$). These patterns are consistent with the weak overlap between the maxima of CAPE and MSE$_s$ (Fig. S2$D$), as well as the low annual percentiles of precipitation following peak MSE$_s$ (Fig. S2$F$), which indicates a less frequent co-occurrence of extreme moist heat and intense convection, and thus processes other than convective precipitation may be responsible for terminating extreme heat in these regions.

\section*{Theoretical basis}

\subsection*{CAPE scaling}

We start with providing a scaling of CAPE in terms of MSE$_s$ to demonstrate why both moist heat and potential convection maximize at the same time. Based on the definition of parcel buoyancy and simple assumptions about the geometry of CAPE, CAPE can be approximated by the difference between MSE$_s$ and MSE$_{500}^{*}$ scaled by a factor of 0.22 (detailed in $Materials$ $and$ $Methods$):
\begin{equation} 
\label{scaling-cape}
CAPE\simeq 0.22(MSE_{s}-MSE^{*}_{500}).
\end{equation} 

The scaling CAPE (right hand side of Eq.\ref{scaling-cape}) builds a simple linear relation between CAPE and basic environmental parameters of the near-surface and free tropospheric atmospheres, without relying on the profiles of a hypothetically lifted air parcel. The scaling CAPE$_c$, estimated using MSE$_s$ and MSE$^{*}_{500}$ at the time of annual maximum MSE$_s$, effectively captures the spatial distribution of CAPE$_c$ (pattern correlation coefficient = 0.98; Fig. \ref{fig_02}$A$ and $B$), with values closely aligned along the one-to-one line (Fig. \ref{fig_02}$C$). Note that the virtual temperature correction is considered in our calculation of standard CAPE but ignored in our derivation of scaling CAPE for simplicity, but the strong agreement between the two suggests that virtual effects are minor. Since CAPE$_c$ is generally high, this is consistent with previous findings that virtual temperature correction is negligible for large CAPE values \cite{Doswell_Rasmussen_1994}.

If MSE$^{*}_{500}$ does not change much during the heatwave evolution and convection build-up periods (consistent with results found by previous studies \cite{neal_etal_2022,zhang_boos_2023,Tuckman_etal_2023,Tuckman_Emanuel_2024} and supported by Fig. \ref{fig_03}$C$ discussed later), then from Eq.\ref{scaling-cape}, CAPE and MSE$_s$ are expected to reach their peak intensities roughly at the same time. The frequent concurrence of extreme near-surface moist heat and convective instability suggests a shared governing process that drives the accumulation of both phenomena over midlatitude land, thereby the factor determining the maximum MSE$_s$ simultaneously sets the maximum CAPE. We proceed to provide a theoretical prediction for both maximum moist heat and maximum potential convection.

\subsection*{A theory for convection onset}

We stick to the hypothesis that the onset of deep convection sets the maximum potential intensity of both near-surface moist heat and convection, but we incorporate a more detailed consideration of convection onset to propose a new theoretical framework. This framework allows surface heat and convective instability to evolve beyond the neutral condition, providing a tight constraint on both near-surface moist heat and convection throughout the process.

\begin{figure*}[tbhp]
\centering
\centerline{\includegraphics[width=0.9\linewidth]{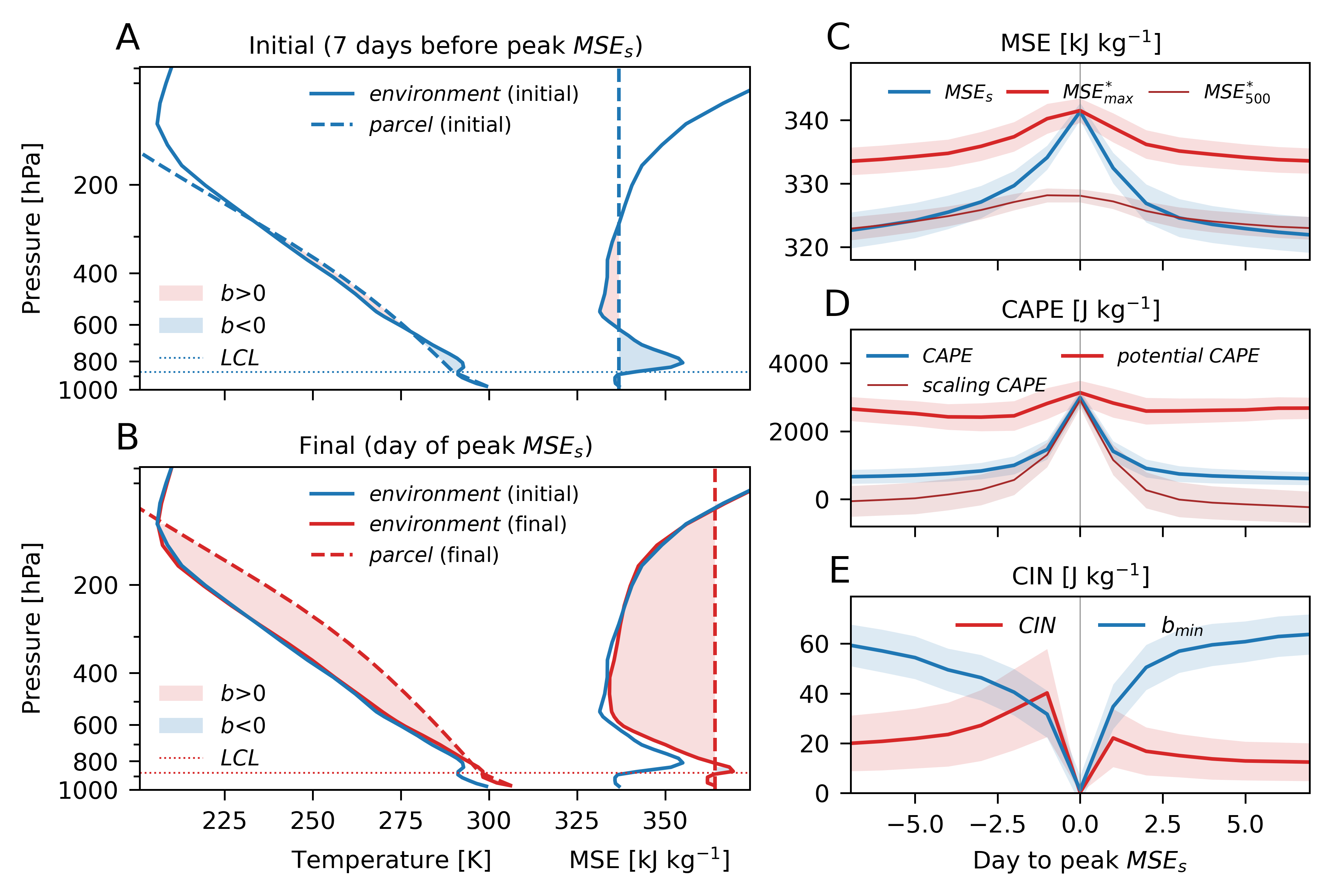}} 
\caption{Moist heat and convection buildup. ($A$--$B$): A case study associated with the annual maximum $MSE_s$ over central United States (41$^{\circ}$N, -96$^{\circ}$W) in 2004, temperature sounding (left) and MSE (right) profiles for the environment (solid lines) and the air parcel (dashed lines) adiabatically lifted from the near surface, at the time of ($A$) 5 days before the annual maximum $MSE_s$ and ($B$) the maximum $MSE_s$ during that year. Shaded areas indicate positive ($b>0$) and negative ($b<0$) buoyancy, and the dotted lines represent lifted condensation level ($LCL$). ($C$--$E$): Composite time series, centered at the time of annual maximum $MSE_s$, for ($C$) $MSE_s$, $MSE_{max}^{*}$, and $MSE_{500}^{*}$; ($D$) $CAPE$, scaling $CAPE$, and potential $CAPE$; ($E$) $CIN$ and the most negative buoyancy ($b_{min}$, scaled by 0.005). Shading denotes $\pm$ one standard error of the mean. The ($C$--$E$) composites are based on midlatitude continental cases where the annual maximum $MSE_s$ and $CAPE$ occur concurrently, with $CAPE_c>500$ J kg$^{-1}$, $CIN_c<5$ J kg$^{-1}$, and the maximum $CIN$ within 7 days before the peak heat of at least 50 J kg$^{-1}$ (detailed in $Materials$ $and$ $Methods$). Results are based on ERA5 reanalysis data. The ERA5 profiles for the case presented in ($A$--$B$) are validated against radiosonde observations form the National Weather Service in Omaha, NE, USA, as shown in Fig. S3. 
}
\label{fig_03}
\end{figure*}

The foundation of our framework is grounded in the stored-energy nature of severe convection over midlatitude continents \cite{doswell_1987,Emanuel_1994,Doswell_2001,chaboureau_etal_2004,bechtold_etal_2004,Agard_Emanuel_2017,Emanuel_2023,lafleur_etal_2023}, primarily attributed to a preexisting stable lower free troposphere \cite{Carlson_etal_1983,colby_1984,Ribeiro_Bosart_2018,raymond_etal_2021,Li_etal_2020,Li_etal_2021,Tuckman_etal_2023,Tuckman_Emanuel_2024,andrews_etal_2024}. For instance, as shown in a sample sounding from 5 days prior to the annual maximum surface moist heat in 2004 over the central United States (temperature profiles in Fig. \ref{fig_03}$A$), the low-level atmospheric stability typically creates a strong layer of negative buoyancy (i.e., $b$ < 0 and CIN > 0; initial CIN=180 J kg$^{-1}$ for the case) that inhibits convective initiation, even when the free troposphere may already be unstable (i.e., $b$ > 0 and CAPE > 0; initial CAPE=672 J kg$^{-1}$ for the case). Therefore, assuming sufficient heating sources, the near-surface atmosphere can continue to warm or moisten until the negative buoyancy layer is eroded, enabling the initiation of deep free convection to release CAPE and terminate the surface heat through precipitation followed by evaporative cooling. In the case study presented, near-surface heat progressively increased over the following days, accompanied by a rise in CAPE. Meanwhile, the free troposphere remained relatively constant until the time of maximum surface moist heat, when convective inhibition was nearly eliminated (temperature profiles in Fig. \ref{fig_03}$B$). At the critical point, when both the MSE$_s$ and CAPE reach their maximum values, the local atmospheric environment is characterized by CIN$_c\simeq$ 0 and CAPE$_c$ $\gg$ 0 (temperature profiles in Fig. \ref{fig_03}$B$; final CIN=9 J kg$^{-1}$ and CAPE=6753 J kg$^{-1}$ for the case), consistent with the results presented above (Fig. S2$A$ and \ref{fig_02}$A$). 

Therefore, the regime of near-surface moist heat accumulation and convection buildup is constrained by CIN $\leq$ 0. Since CIN essentially reflects the presence of negative buoyancy layers within the lower atmosphere, whether and when CIN approaches 0 is governed by the evolution of the most negative buoyancy ($b_{min}(z)$) throughout the low levels \cite{Tuckman_etal_2023,Tuckman_Emanuel_2024}. Therefore, CIN $\leq$ 0 is effectively equivalent to $b_{min}(z)$ $\leq$ 0. The latter is more general and practical, as it also captures cases where CIN is 0 by definition due to the absence of a defined level of free convection (LFC) and equilibrium level (EL), with the entire atmospheric column characterized by negative buoyancy for the near-surface lifted air parcel. 

As parcel buoyancy is proportional to the (virtual) temperature difference between the lifted air parcel and its environment, we further utilize the conservation of the parcel's MSE to derive an alternative formulation of temperature perturbations entirely as a function of environmental static energies (Eq.8, detailed in $Materials$ $and$ $Methods$). This also provides an MSE perspective to visualize the profiles of temperature perturbations and buoyancy by comparing the parcel's initial MSE (i.e., $MSE_s$) with the environmental moist static energy profile, which is $DSE(z)+L_vq_s$ for $z< LCL$ or $MSE^{*}(z)$ for $z\geq LCL$, where $LCL$ represents the lifted condensation level (Eq.9, detailed in $Materials$ $and$ $Methods$). As $b_{min}(z)$ is usually above the $LCL$ \cite{Tuckman_etal_2023,Tuckman_Emanuel_2024}, based on Eq.9, $b_{min}(z)$ $\leq$ 0 is equivalent to

\begin{equation}  
\label{eq1.1}
b_{min}(z)\sim min\{MSE_s-MSE^*(z)\}\leq 0
\end{equation}

As MSE$_s$ does not change with height, $min\{MSE_s-MSE^*(z)\}$ = MSE$_s-$MSE$_{max}^{*}$, where MSE$_{max}^{*}$ is the maximum MSE$^*$ over the lower free tropospheric layers above or at LCL (detailed in $Materials$ $and$ $Methods$). The MSE$_{max}^{*}$ quantifies the strongest energy barrier within the lower free troposphere that a near-surface air parcel must overcome to initiate convection in the absence of external forcing. Therefore, the maximum $MSE_s$ is limited by MSE$_{max}^{*}$, such that
\begin{equation}  
\label{eq1.2}
MSE_s\leq MSE^*_{max}
\end{equation}

In addition, from the framework of scaling CAPE (Eq.\ref{scaling-cape}), the MSE$_{max}^{*}$ and MSE$_{500}^{*}$ together constrain CAPE$_c$ by
\begin{equation}  
\label{eq1.3}
CAPE_c\leq 0.22(MSE^*_{max}-MSE^*_{500})
\end{equation}

The right-hand side of Eqs.\ref{eq1.2} and \ref{eq1.3} defines the maximum potential intensities of near-surface moist heat (MSE$_s$) and convection (CAPE), respectively. These maximum potential intensities are achievable under conditions where surface heat sources are sufficient and external lifting forces are absent, requiring $b_{min}(z)$ to approach zero to enable free convection, which is commonly observed in midlatitude continents \cite{colby_1984,Rasmussen_Blanchard_1998,Taszarek_etal_2020,Tuckman_etal_2023,Tuckman_Emanuel_2024}. Otherwise, the actual intensities are likely to be lower than their maximum potential values.

\section*{Maximum potential moist heat and convection}

We first present the results of the sounding evolutions for the case discussed above, analyzed within the developed MSE framework. Those profiles, taken from the ERA5 reanalysis data, closely align with radiosonde observations (Fig. S3). As a validation of Eq.9, the MSE profiles (Fig. \ref{fig_03}$A$ and $B$) recover well the positive and negative buoyancy layers and their temporal evolutions. Compared to temperature profiles, the MSE profiles more clearly delineate the most negative buoyancy layer prior to the annual maximum moist heat, identified by a distinct MSE$^*$ inversion above the LCL (Fig. \ref{fig_03}$A$). The preexisting MSE$^*_{max}$ resides at the peak of the MSE$^*$ inversion near 800 hPa (MSE profiles in Fig. \ref{fig_03}$A$), and eventually aligns with the LCL (MSE profiles in Fig. \ref{fig_03}$B$). In this case, the final MSE$^*_{max}$ effectively sets a constraint that is only slightly greater than the maximum MSE$_s$ achieved at the time of the annual maximum moist heat, when $b_{min}$ (and hence CIN$_c$) approaches nearly 0 and CAPE$_c$ is maximized (CIN$_c$=9 J kg$^{-1}$ and CAPE$_c$=6753 J kg$^{-1}$). The small departure of peak MSE$_s$ form the final MSE$^*_{max}$ is consistent with the small residual of CIN$_c$. Relative to the significant changes in MSE$_s$ during this period, the preexisting MSE$^*_{max}$ intensities weakly over time, which suggests the possibility of predicting the maximum potential intensity of MSE$_s$ (and consequently CAPE$_c$) based on the initial MSE$^*_{max}$, although adjustments are needed over time as MSE$^*_{max}$ evolves. 

\begin{figure*}[tbhp]
\centering
\centerline{\includegraphics[width=1.03\linewidth]{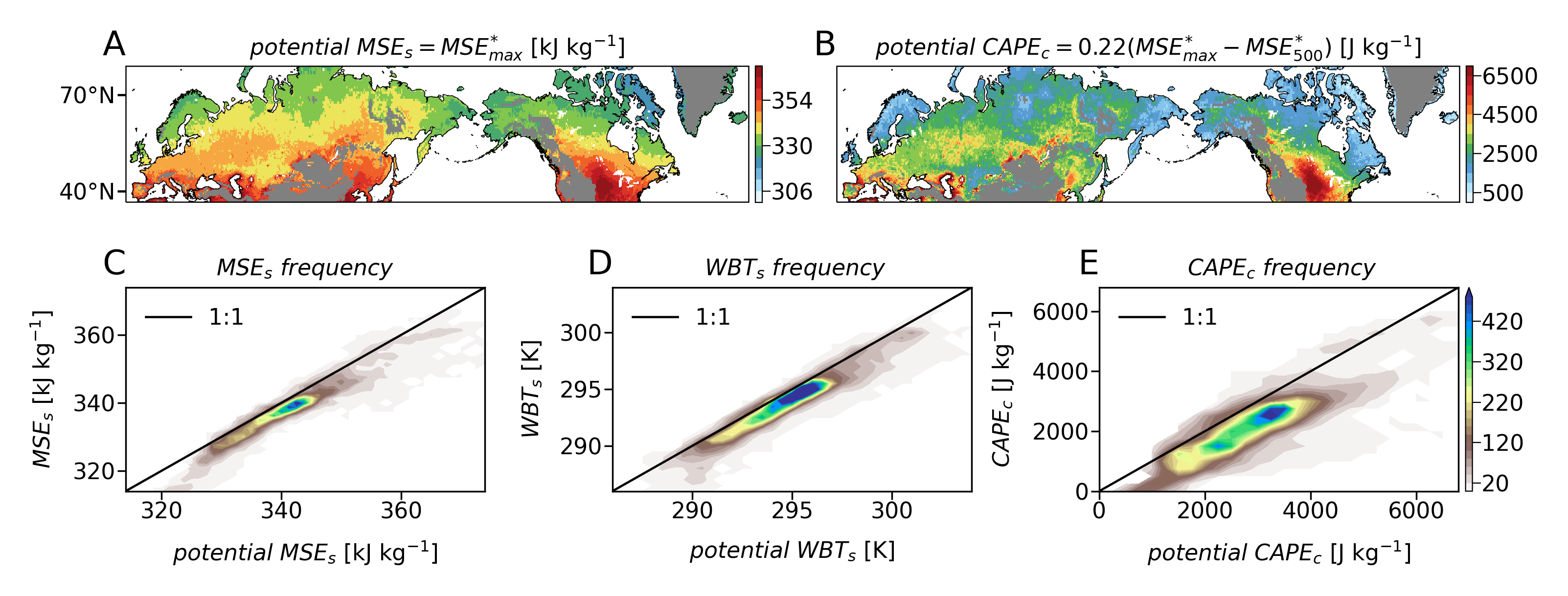}}
\caption{Maximum potential moist heat and convection. ($A$): Annual maximum potential $MSE_s$, defined as $MSE_{max}^{*}$ corresponding to the maximum $MSE_s$ following Eq.\ref{eq1.2}. ($B$): Annual maximum potential $CAPE_c$, defined by CAPE scaling based on $MSE_{max}^{*}$ and $MSE_{500}^{*}$ following Eq.\ref{eq1.3}. ($C$--$E$): Joint histograms of ($C$) annual maximum $MSE_s$ and the potential intensity with a bin size of 1.25$\times$1.25 kJ kg$^{-1}$, ($D$) annual maximum $WBT_s$ and the potential intensity with a bin size of 0.5$\times$0.5 K, and ($E$) $CAPE_c$ and the potential intensity with a bin size of 250$\times$250 J kg$^{-1}$. Results are historical means based on ERA5 reanalysis data during 1980--2022 for land between 35$^{\circ}$N and 75$^{\circ}$N at elevations lower than 1000 m.} 
\label{fig_04}
\end{figure*}

To evaluate the generality of the above results, we extend the analysis by presenting composite time series of key proxies derived from a subset of cases over midlatitude land. Specifically, we focus on instances of concurrent annual maxima in MSE$_s$ and CAPE for each year during 1980--2022 (44.5\% of all cases; Fig. S2$D$). To best visualize the features aligning with our theoretical framework - where MSE$_s$ and CAPE are maximized as preexisting CIN is eroded to zero - we further select cases based on the following criteria: CAPE$_c$ > 500 J kg$^{-1}$, CIN$_c\leq$ 0, and a maximum CIN of at least 50 J kg$^{-1}$ within seven days preceding the critical time (detailed in $Materials$ $and$ $Methods$). These selection criteria yield 2502 cases per year (27\% of concurrent cases), for which we plot composite 14-day time series centered on the time of maximum MSE$_s$ (Fig. \ref{fig_03}$C$–$E$). Following the maximum potential for moist heat defined by Eq.\ref{eq1.2}, MSE$_s$ increases over time and ultimately converges toward MSE$^*_{max}$ at the peak time (Fig. \ref{fig_03}$C$). The height of MSE$^*_{max}$ is in general between 750--800 hPa, and decreases gradually with time to approach $LCL$ (Fig. S4$B$). Notably, MSE$^*_{500}$ fails to predict the maximum MSE$_s$ as discussed above, as MSE$_s$ exceeds MSE$^*_{500}$ as early as seven days before the peak, with the difference rapidly amplifying over time (Fig. \ref{fig_03}$C$). This growing discrepancy between MSE$_s$ and MSE$^*_{500}$ reflects the progressive accumulation of CAPE leading up to the peak, as recovered by the scaling CAPE following Eq.\ref{scaling-cape} and constrained by the maximum potential convection defined in Eq.\ref{eq1.3} (Fig. \ref{fig_03}$D$). The peak CAPE on the hottest MSE$_s$ day, followed by its decrease, indicates the onset of convection that consumes convective instability and terminates extreme heat, which is further reflected by daily precipitation that also peaks on the hottest day and remains relatively high over the following days (Fig. S4$A$). Fundamentally, the composite temporal evolutions of MSE$s$ and CAPE are linked to the removal of CIN or $b_{min}$ (Fig. \ref{fig_03}$E$), supporting our basic hypothesis formulated in Eq.\ref{eq1.1}. These results are broadly consistent when we perform composite analyses based on all cases of concurrent moist heat and convection over midlatitude land (detailed in $Materials$ $and$ $Methods$; Fig. S5), though the maximum potential intensities of MSE$_s$ and CAPE$_c$ are not necessarily achieved (Fig. S5$A$--$B$). This suggests the potential influence of external forcing mechanisms of lift \cite{weckwerth_Parsons_2006,marsham_etal_2011,kuo_Neelin_2022}, which may facilitate the initiation of convection in cases with non-zero CIN$_c$ (Fig. S5$C$). Similarly, the composite results reveal relatively steady time series for MSE$^*_{max}$ and MSE$^*_{500}$ (Fig. \ref{fig_03}$C$), as well as for the potential CAPE (Fig. \ref{fig_03}$D$), highlighting the potential to predict the maximum intensities of MSE$_s$ and CAPE based on preexisting free tropospheric properties.    

Lastly, to comprehensively validate the maximum potential intensity framework (i.e., Eqs.\ref{eq1.2} and \ref{eq1.3}), we compare the observed annual maximum moist heat (MSE$_s$) and the associated critical convection (CAPE$_c$) with their potential maxima derived for all cases over midlatitude land during 1980--2022 from the ERA5 reanalysis data (Fig. \ref{fig_04}). Here, MSE$^*_{max}$ and MSE$^*_{500}$ are calculated at the time of annual maximum MSE$_s$. Overall, the maximum potential MSE$_s$ (Fig. \ref{fig_04}$A$) and the maximum potential CAPE$_c$ (Fig. \ref{fig_04}$B$) capture the spatial patterns of the observed maximum MSE$_s$ (Fig. \ref{fig_01}$B$) and CAPE$_c$ (Fig. \ref{fig_02}$A$) over midlatitude land, with observed maxima slightly smaller than their potential counterparts. The pattern correlation coefficient is 0.94 between observed and potential MSE$_s$ maxima, and 0.82 between observed and potential CAPE$_c$. The joint histograms of all cases further confirm that the maximum potential intensities serve as a tight upper limit for the observed maximum MSE$_s$ and CAPE$_c$ (Fig. \ref{fig_04}$C$ and $E$). For the maximum MSE$_s$, the potential MSE$_s$ is achievable for many cases, with others closely aligned with the one-to-one line but slightly below (Fig. \ref{fig_04}$C$). Translating (potential) MSE$_s$ into (potential) WBT$_s$ (as detailed in Materials and Methods) yields consistent results (Fig. \ref{fig_04}$D$), highlighting the applicability of our theory for directly predicting maximum WBT$_s$. Similarly, the potential CAPE$_c$ serves as a strong constraint on the observed CAPE$_c$ (Fig. \ref{fig_04}$E$), although the actual CAPE$_c$ exhibits greater variability for a given potential CAPE$_c$, partly due to the variations in MSE$_s$. It is worth noting the relatively large discrepancy over the land east of the Caspian Sea, south of the Mediterranean Sea, and along the west coast of the United States (Fig. \ref{fig_04}$A$--$B$ versus Fig. \ref{fig_01}$B$ and \ref{fig_02}$A$), where the residual energy barrier (i.e., the difference between MSE$^*_{max}$ and MSE$_s$ at the peak time) is high, consistent with the high CIN$_c$ discussed above. This again suggests that processes other than convection are responsible for terminating extreme heat events before surface heating and convection can approach their potential intensities in regions where convective inhibition persists.  

\section*{Conclusion and discussion}

This study proposed a theoretical prediction for maximum potential moist heat (MSE$_s$) and convection (CAPE) over midlatitude land, both tightly constrained by a preexisting energy (MSE$^*$) inversion in the lower free troposphere. The maximum MSE$^*$ at the inversion layer (MSE$^*_{max}$) acts as a substantial energy barrier that suppresses the free convection of near-surface air parcels, thereby allowing near-surface warming or moistening and the accumulation of convective instability. This process can continue until MSE$_s$ reaches MSE$^*_{max}$, such that the near-surface air parcels are sufficiently energetic to overcome the energy barrier, initiate deep free convection, and terminate surface heating. This strong buildup of CAPE, which enables excessive increases in surface moist heat, reflects the stored-energy signature of severe convection in the midlatitudes \cite{doswell_1987,Emanuel_1994,Doswell_2001,chaboureau_etal_2004,bechtold_etal_2004,Zipser_etal_2006,Agard_Emanuel_2017,Emanuel_2023,lafleur_etal_2023,Tuckman_etal_2023,Tuckman_Emanuel_2024}. 

Our theory advances previous work on convective constraints of extreme heat by explicitly incorporating CAPE buildup such that the atmosphere can evolve beyond the neutral condition (i.e., parcel buoyancy and CAPE are zero by definition), and thus extends many previous studies commonly assuming moist neutrality at the peak time \cite{buzan_huber_2020,zhang_etal_2021_heat,byrne_2021,zhang_boos_2023}. The buildup of CAPE is a non-negligible factor in predicting surface heat extremes: accumulation of CAPE from 0 to 1000 J kg$^{-1}$ corresponds an increase of $\sim$4.5 K in surface temperatures ($\Delta T_s\approx \frac{1}{0.22c_p}\Delta CAPE=4.52 K$ from Eq. \ref{scaling-cape}), assuming other conditions remain constant. One prior study proposed that surface moist heat is limited by the tropopause MSE$^*$ \cite{buzan_huber_2020}, which aligns with our prediction when the low-level inversion barrier is as strong as the tropopause MSE$^*$. However, not all convection reaches the tropopause in the midlatitudes, and the tropopause constraint tends to overestimate the maximum intensity of surface moist heat if the low-level inversion energy is weaker than the tropopause MSE$^*$. In addition, a recent study examined extreme dry heat over midlatitude continents under the assumption of moist neutrality \cite{zhang_boos_2023}, suggesting that 500-hPa MSE$^*_{500}$ acts as an upper bound for MSE$_s$. However, we have shown that the neutrality assumption fails for extreme moist heat over midlatitude land (Fig. \ref{fig_01}$D$). In fact, the neutrality assumption also breaks down for extreme dry heat in the midlatitudes, as MSE$_s$ exceeds MSE$^*_{500}$ in most cases of annual maximum surface air temperature (Fig. S6$A$). Our CAPE scaling framework (Eq.\ref{scaling-cape}) reveals that MSE$^*_{500}$ generally approximates the minimum energy of the free troposphere (detailed in $Materials$ $and$ $Methods$), and the surplus of MSE$_s$ over MSE$^*_{500}$ effectively defines the accumulated convective instability beyond moist neutrality. The dry limit theory \cite{zhang_boos_2023} further assumes completely dry surface conditions that limit the accumulation of moist convective instability and result in an unrealistically high upper bound on surface temperature constrained by a deep dry adiabatic layer. However, this drying assumption contradicts the hypothesis that convective precipitation terminates heatwaves, as moisture accumulation is essential for the initiation and development of moist convection. In contrast, our theory does not require assumptions about surface humidity and explicitly allows the buildup of moist convective energy, therefore offering a tighter and more physically grounded constraint on surface heat. When applied to extreme dry heat, our theory also predicts well the maximum potential intensities of MSE$_s$ and CAPE$_c$ (Fig. S6$B$–$C$), which further generalizes the theory. 

Within the context of convective constraints on heat extremes, accurately identifying the onset of convection is crucial for determining both the termination and peak intensity of surface heat. Future work could incorporate other processes (e.g., wind shear \cite{peters_etal_2022_shear1,peters_etal_2022_shear2,moncrieff_liu_1999}, entrainments \cite{singh_ogorman_2013,peters_etal_2023,duan_etal_2024}, and external forcings \cite{rasmussen2016,nelson_etal_2022,moncrieff_liu_1999,banacos_Schultz_2005,Ziegler_etal_1997,bennett_etal_2006review,kuo_Neelin_2022,weckwerth_Parsons_2006,marsham_etal_2011}) to more precisely predict convection onset and, consequently, the termination of heatwaves. For example, a recent study emphasized the role of dry air entrainment in suppressing deep convection and intensifying moist heat in tropical and subtropical regions \cite{duan_etal_2024}, which is quantified by the production of entraining rate ($\omega=1$) and 850-hPa saturation deficit (MSE$_{850}$-MSE$^*_{850}$). Incorporating sub-cloud entrainment into our framework could further modify the convective constraint on surface moist heat as approximately MSE$^*_{max}$+$\omega$(MSE$^*_{850}$-MSE$_{850}$). While this may increases the weight of low-level MSE$^*$, the entraining factor $\omega$ needs careful estimation in the midlatitudes, where vertical wind shear associated with convection is typically stronger than in the tropics \cite{Brooks_etal_2003}. In midlatitude severe convection, the strong deep-layer vertical wind shear tends to broaden convection and prevent the influence of entrainment (i.e., smaller $\omega$ under larger deep-layer wind shear) \cite{mulholland_etal_2021,mulholland_etal_2024,peters_etal_2024}.

\section*{Implication and outlook}

Our work centered on parcel theory to quantify the upper bound on local near-surface moist heat and convection before local free convection is triggered. In most midlatitude land regions, extreme moist heat and convection occur concurrently, and MSE$^*_{max}$ provides a tight and attainable constraint on their maximum intensities. In certain regions (e.g., east of the Caspian Sea, south of the Mediterranean, and along the west coast of the United States), this energetic inversion barrier (MSE$^*_{max}$) remains largely unbroken, and convective precipitation appears incapable to terminate extreme heat events. In these areas, moist heat and convection extremes co-occur less frequently, and processes other than convective precipitation may terminate extreme heat events, which deserves further future investigation. While our theory does not address the specific mechanisms by which surface moist heat or convective instability increases, previous studies have examined a range of contributing processes, including warm air or moisture advection, adiabatic warming in subsiding air, and diabatic heating from surface fluxes and radiation, associated with large-scale dynamics such as storm tracks and atmospheric blocking \cite{pfahl_Wernli_2012,bieli_etal_2015,rothlisberger_Papritz_2023,Tuckman_Emanuel_2024,mckinnon_etal_2024,tamarin_etal_2020,kong_huber_2023,fischer_etal_2007,miralles_etal_2014,Molina_Allen_2019,neal_etal_2022,li_Mann_etal_2024}. Bridging these dynamical drivers with thermodynamical constraints is essential for predicting and understanding how, when, and to what extent heatwave and severe convection can approach theoretical limits, and why such limits are not reached in some regions or events. 

Our theory emphasizes the critical role of low-level energy inversion layers in shaping extreme moist heat and severe convection events. These inversion features typically form several days before the peak events and vary weakly over time, therefore offering a valuable opportunity for predicting the maximum intensity of moist heat and convection on synoptic timescales. This predictive potential has long been recognized by Severe Local Storm research and operational communities. A notable example is the elevated mixed layer, characterized by a pronounced low-level temperature inversion formed by warm air advection from upstream elevated terrain, which has been identified as a common synoptic-scale feature contributing to intense severe thunderstorm and tornado activity across central North America \cite{Carlson_etal_1983,Banacos_Ekster_2010,Li_etal_2020,Li_etal_2021,andrews_etal_2024}, South America \cite{Ribeiro_Bosart_2018}, and Europe \cite{schultz_2025spanish}. Recent studies using Lagrangian trajectory methods and analytical models further suggest that such inversion features are widespread and play a key role in many severe convective events globally \cite{Agard_Emanuel_2017,Tuckman_etal_2023,Emanuel_2023,Tuckman_Emanuel_2024}. Our work quantifies the energetic constraint imposed by inversion layers and extends its utility to extreme moist heat. In practice, monitoring the presence of an explicit low-level MSE$^*$ inversion could serve as a key signal for timing predictions of the potential intensities of surface moist heat and convection extremes. Meanwhile, understanding the evolutions of MSE$^*_{max}$ and MSE$^*_{500}$ is also crucial for refining these predictions over time. Since both heatwaves and severe convective storms are among the most damaging weather extremes, this theory represents a promising step toward improving the prediction and understanding of these events and their potential compound occurrence in the midlatitudes.

Lastly, our theoretical framework provides a valuable lens for predicting and understanding extreme heat and severe convection under future climates. While low-level energy inversion layers serves as local convective barriers, their formations may stem from non-local processes, such as warm-air advection from an adjacent region (e.g., elevated terrain) or adiabatic warming from subsiding air within large-scale circulations (e.g., anticyclones) \cite{kassomenos_etal_2014,Tuckman_Emanuel_2024,Carlson_etal_1983,Li_etal_2021,raymond_etal_2021}. A Lagrangian perspective can help trace the origins and transport pathways of inversions \cite{tamarin_etal_2020,tamarin_Kaspi_2017,keune_etal_2022,Tuckman_etal_2023,rothlisberger_Papritz_2023,Tuckman_Emanuel_2024}, offering insight into MSE$^*_{max}$ evolution over time. As the climate changes, land cover change and uneven regional warming may alter inversion formation at the source \cite{mueller_etal_2016,barnes_etal_2024}, while changes in large-scale driven dynamics (e.g., elevation-dependent warming and storm track shifts) may influence airmass transport pathways \cite{mountain_2015,byrne_etal_2024,tamarin_Kaspi_2017,tamarin_Kaspi_2016,tamarin_Kaspi_2017_grl,shaw_Miyawaki_2024}. These shifts could affect the intensity, height, persistence and spatial extent of inversion layers \cite{andrews_etal_2024,Emanuel_2023}, thereby reshaping future spatial patterns of extreme heat and convective weather \cite{tamarin_etal_2019,vecellio_etal_2023,tang_etal_2019,jiang_etal_2025}. For example, elevated temperature inversions have intensified over the central United States \cite{andrews_etal_2024}, likely associated with the amplified warming upstream over the Rocky Mountains \cite{mountain_2015,byrne_etal_2024}, which may raise low-level energy barriers and imply that greater near-surface heating will be required to trigger convection. These changes may be responsible for the amplified moist heat stress over central and Midwest United States\cite{vecellio_etal_2023} and the eastward expansion of severe thunderstorms in this region \cite{jiang_etal_2025,tang_etal_2019}. Climate change may also alter external forcings (e.g., fronts, cyclogenesis, moisture convergence) \cite{catto_etal_2014,catto_etal_2015}, which can modify the timing of convection onset and drive spatial variations in the divergence between realized and potential intensities of heat and convection extremes. Finally, our framework underscores the need to evaluate model skill in simulating low-level inversions and convection initiation, as the associated biases can substantially impact projections of extreme heat, severe convection, and their compound occurrence \cite{lin2017,chavas_li_2022_biases}.

\matmethods{

\subsection*{Data}

We use 3-hourly near-surface and pressure-level ERA5 reanalysis data with a horizontal grid spacing of 0.5$^\circ$ latitude by 0.5$^\circ$ longitude during 1980--2022 \cite{NCAR_RDA_ERA5,Hersbach_etal_2020}, for land grid points between 35$^{\circ}$N and 75$^{\circ}$N at elevations lower than 1000 m (20854 grid points per year in total). ERA5 reanalysis data are widely used in studies of global temperature extremes and convective environments, as they accurately capture observed temperature and moisture profiles in general \cite{Taszarek_etal_2021,Taszarek_etal_2021_global,Tuckman_Emanuel_2024}, including inversion features within the lower free troposphere \cite{Li_etal_2020, andrews_etal_2024}. In this study, calculations for all variables are initially performed at the 3-hourly interval and then sampled based on the daily maximum MSE$_s$ to create a dataset of daily maximum moist heat. Our analyses focus on the annual maximum moist heat, obtained by further sampling the dataset given the annual maximum MSE$_s$ at each grid point for each year. This process yields 43 samples per land grid point, corresponding to the maximum MSE$_s$ for each year during 1980--2022. For analyses related to extreme near-surface dry heat, the data are generated in the same manner but are conditioned on the maximum near-surface dry heat, defined as the maximum 2-m air temperature (T$_s$). 

\subsection*{MSE and WBT}

MSE is defined as the sum of sensible heat ($c_pT$), latent heat ($L_vq$), and geopotential energy ($gz$), i.e., $MSE=c_pT+L_vq+gz$, where $c_{p}=1005$ J kg$^{-1}$ K$^{-1}$ is the the specific heat capacity of air at constant pressure, $L_{v}=2.5\times 10^6$ J kg$^{-1}$ is the latent heat of vaporization, $g=9.81$ m s$^{-2}$ is gravitational acceleration, $T$ is air temperature, $q$ is specific humidity, and $z$ is height above sea level \cite{Emanuel_1994}. Near-surface (denoted by subscript ``$s$'') MSE, $MSE_s$, is calculated using quantities at 2-m above ground surface, where the geopotential height equals to 2 plus the surface elevation in meters. Saturated (denoted by superscript ``$*$'') MSE, $MSE^*$, is calculated by replacing $q$ with the saturated specific humidity ($q^*$), which is a function of air temperature and pressure. In an adiabatic process, MSE is generally considered a conserved quantity given hydrostatic balance \cite{Emanuel_1994,romps2015,peters_etal_2021_conservation}, interconvertible with equivalent potential temperature \cite{madden_Robitaille_1970,betts_1974,chavas_peters_2023}. As a part of MSE, dry static energy (DSE) is the sum of  $c_pT$ and $gz$, which is conserved under dry adiabatic conditions and is thermodynamically equivalent to potential temperature \cite{Emanuel_1994,decaria_2007}. $MSE_{max}^*$ is identified as the maximum MSE$^*$ within the lower free troposphere, specifically at or above $LCL$ but below 300 hPa. 

$WBT_s$ is calculated based on the conservation of $MSE_s$ \cite{eltahir_pal_1996,zhang_etal_2021_heat}. Specifically, for each $MSE_s$, we solve the equation $MSE_s=c_pWBT_s+L_vq^*(WBT_s,p_s)+gz_s$ for the corresponding $WBT_s$, where $q^*(WBT_s,p_s)$ is the saturated specific humidity at temperature $WBT_s$ and surface pressure $p_s$. As $q^*(WBT_s,p_s)$ is proportional with $WBT_s$ based on the Clausius–Clapeyron scaling \cite{ogorman_Muller_2010}, $WBT_s$ exhibits a nearly one-to-one relationship with $MSE_s$ (Fig. S1) \cite{zhang_etal_2021_heat,raymond_etal_2021}. There is a small variation in WBT$_s$ given a value of MSE$_s$, due to dependence on the elevation variance (0--1000 m; Fig. S1). Similarly, we calculate the maximum potential $WBT_s$ (Fig. \ref{fig_04}$D$) by solving $WBT_s$ from the equation $MSE_{max}^*=c_pWBT_s+L_vq^*(WBT_s,p_s)+gz_s$ at the time of annual maximum $MSE_s$.

\subsection*{Composite time series}

Composite time series analyses (Fig. \ref{fig_03}$C$--$E$) are based on a subset of extreme moist heat and convection cases. To focus on the locally most extreme moist cases, we first select cases of annual maximum $MSE_s$ when its CAPE$_c$ corresponds to the local annual maximum CAPE (i.e., cases with moist heat and convection both maximized at the same time during the year). This step on average yields 9270 cases per year (i.e., 44.5\% of all cases), with mean CAPE of up to 3011 J kg$^{-1}$ and mean CIN as low as 12 J kg$^{-1}$. There are a few cases over far northern Canada (Nunavut area), where convection is rare and the annual maximum CAPE is less than 500 J kg$^{-1}$. Thus, those cases with CAPE$_c\leq$ 500 J kg$^{-1}$ are excluded, which further yields 8959 cases per year. Composite time series based on this large group of cases is shown in Fig. S5. The composite time series indicates a substantial increase of MSE$_s$ bounded by MSE$^*_{max}$ (Fig. S5$A$) and the buildup of CAPE bounded by the potential CAPE (Fig. S5$B$), along with the removal of the most negative buoyancy (Fig. S5$C$). These results are broadly consistent with that shown in main text (Fig. \ref{fig_03}$C$--$E$) .

We further select cases where the theoretical maximum potential intensities (Eqs.\ref{eq1.2}--\ref{eq1.3}) are likely achieved and the preexisting energy barrier is relatively large, by conditioning on CIN$_c\leq$ 5 J kg$^{-1}$ and the maximum CIN of at least 50 J kg$^{-1}$ within 7 days preceding the critical time. This leads to 2502 cases per year (i.e., 12\% of all cases), with the composite time series shown in the main text (Fig. \ref{fig_03}$C$--$E$).  

\subsection*{Parcel buoyancy and temperature perturbation}

To link surface moist heat with convection, we begin by deriving an expression for parcel buoyancy as a function of near-surface moist static energy.

Considering an undiluted air parcel adiabatically lifted upward from the near-surface (2 m above ground surface; denoted by subscript ``$s$'' in derivations) and neglecting the relatively small contribution to density changes due to pressure perturbation and virtual temperature correction\cite{Doswell_Rasmussen_1994, Emanuel_1994}, the parcel buoyancy at a given height $z$ ($b(z)$) is proportional to differences in sensible heat between the air parcel and the environment ($c_p\Delta T(z)$), where $\Delta T(z)=T_a(z)-T(z)$ is temperature perturbations of the air parcel ($T_a(z)$, with subscript ``$a$'' referring to the air parcel) with respect to the environment ($T(z)$) and $c_{p}=1005$ J kg$^{-1}$ K$^{-1}$ is the the specific heat capacity of air at constant pressure. During an adiabatic process, it is common to assume that the air parcel conserves its moist static energy such that $MSE_a(z)=MSE_s$. Therefore, the vertical profile of $b(z)$ or $c_p\Delta T(z)$ is given by

\begin{equation}  
\label{eq1}
b(z)\sim c_{p}\Delta T(z)= MSE_{s}-DSE(z)-L_{v}q_{a}(z)
\end{equation}
where $DSE(z)$ is vertical profile of the environmental dry static energy. 

To derive parcel buoyancy solely as a function of environmental parameters, independent of the lifted air parcel's profiles, the specific humidity of the air parcel ($q_a(z)$) in Eq.\ref{eq1} can be further approximated using environmental parameters that vary with height ($z$) relative to the lifted condensation level (LCL).  

For $z<LCL$, the air parcel remains unsaturated and maintains its specific humidity at the initial value. Hence, 

\begin{equation}  
\label{eq2}
q_a(z)=q_{s} 
\end{equation}
 
For $z\geq LCL$, the air parcel is saturated, such that the parcel specific humidity equals its saturated specific humidity ($q_a^{*}(z)$, with superscript ``$*$'' referring to a quantity at saturation), which is a function of the parcel temperature and air pressure. As pressure difference between the air parcel and environment is negligible \cite{Doswell_Rasmussen_1994, Emanuel_1994}, the difference in saturated specific humidity between the air parcel and environment is caused by difference in temperature, and thus $q_a^{*}(z)$ can be written as a linear relation to the environmental saturated specific humidity ($q^*(z)$) via $q_a^{*}(z)=q^{*}(z)+(\partial q^{*}(z)/\partial T(z))\Delta T(z)$. We further assume constant $\partial q^{*}(z)/\partial T(z)$ and approximate it by $L_{v}q^{*}(z)/(R_{v}T^{2}(z))$ using Clausius-Clapeyron equation, where $R_v=461$ J kg$^{-1}$ K$^{-1}$ is gas constant for water vapor. Hence,

\begin{equation}  
\label{eq3}
q_a(z)=q_a^{*}(z)\simeq q^{*}(z)+\frac{L_{v}q^{*}(z)}{R_{v}T^{2}(z)}\Delta T(z)
\end{equation}

Substituting Eqs.\ref{eq2} and \ref{eq3} into Eq.\ref{eq1} and rearranging it gives

\begin{equation}
\label{eq4}
b(z)\sim c_p\Delta T(z) \simeq 
\begin{cases} 
DSE_s-DSE(z), &z< LCL\\\\ 

\frac{MSE_{s}-MSE^{*}(z)}{1+\frac{L_{v}^2q^{*}(z)}{c_{p}R_{v}T^{2}(z)}}, &z\geq LCL
\end{cases}
\end{equation}

Eq.\ref{eq4} recovers the vertical profiles of temperature perturbations (Fig. S7$A$). The denominator $1+L_{v}^2q^{*}(z)/(c_{p}R_{v}T^{2}(z))$ is always positive, ranging from $\sim$3 in the lower free troposphere above the LCL to $\sim$1 in the upper free troposphere (Fig. S7$B$), which does not alter the sign of buoyancy across the atmospheric column or the location of the most negative buoyancy within the lower free troposphere (Fig. S7$A$). Therefore, Eq.\ref{eq4} is further simplified in a scaling format,

\begin{equation}
\label{eq5}
b(z)\sim c_p\Delta T(z) \sim 
\begin{cases} 
MSE_{s}-(DSE(z)+L_vq_s), &z< LCL\\\\

MSE_{s}-MSE^{*}(z), &z\geq LCL
\end{cases}
\end{equation}

Eq.\ref{eq5} provides an alternative approach to quantify temperature perturbation profile from the MSE prospective (a more precise form is given by Eq.\ref{eq4}; Fig. S7$A$). This eliminates the need to first calculate the temperature profile of a hypothetically lifted air parcel. Instead, the vertical profile of parcel buoyancy can be determined directly by comparing the parcel's initial MSE (i.e., $MSE_s$) with the environmental static energy profile ($DSE(z)+L_vq_s$ if $z< LCL$ or $MSE^{*}(z)$ if $z\geq LCL$). Furthermore, this approach is equivalent to comparing near-surface potential temperature with the potential temperature profile for $z< LCL$ or the near-surface equivalent potential temperature with the saturation equivalent potential temperature profile for $z\geq LCL$.   

\subsection*{CAPE and scaling CAPE}

We calculate convective available potential energy (CAPE) for the near-surface air parcel by integrating virtual temperature difference between the air parcel ($T_{v,a}$) and environment ($T_v$) with respect to natural logarithm of pressure (ln$p$) from the level of free convection (LFC) to equilibrium level (EL), given by

\begin{equation} 
\label{eq6}
CAPE=-R_d{\int_{p_{LFC}}^{p_{EL}} \Delta T_v dlnp} 
\end{equation}
where $\Delta T_v=T_{v,a}-T_v$, $R_d=287$ J kg$^{-1}$ K$^{-1}$ is the ideal gas constant of dry air, $p_{LFC}$ and $p_{EL}$ are pressure at LFC and EL, respectively. Moist adiabats for the air parcel follow irreversible pseudoadiabatic process \cite{Lepore_etal_2021}. This makes little differences as compared to reversible process \cite{Chen_etal_2020}. Based on ideal gas law, Eq.\ref{eq6} is equivalent to the integral of parcel buoyancy with respect to height. The convective inhibition (CIN) is calculated in the same way as CAPE but for negative buoyancy from the surface to LFC. In this work, CIN is defined as positive by its absolute value. 

Next, we derive a scaling for CAPE from Eq.\ref{eq6} combining simple assumptions. Since $\Delta T_v$ is zero at both LFC and EL and reaches its maximum approximately midway between these levels ($\sim$ 500 hPa, typically above LCL at the height of minimum MSE$^*$; Fig. S4$B$), CAPE can be geometrically approximated as the area of a triangle with $-R_d{\int_{p_{LFC}}^{p_{EL}} dlnp}=R_dln(p_{LFC}/p_{EL}) $ being the base and $\Delta T_v$ at 500 hPa ($\Delta T_{v,500}$, with superscript ``$500$'' referring to a quantity at 500 hPa) being the height. Further neglecting virtual temperature correction (i.e., $\Delta T_{v,500}\simeq \Delta T_{500}$) and substituting Eq.\ref{eq4}, we have that 

\begin{equation} 
\label{eq7}
CAPE\simeq \frac{R_dln(p_{LFC}/p_{EL})}{2c_p}\frac{MSE_{s}-MSE^{*}_{500}}{1+\frac{L_{v}^2q^{*}_{500}}{c_{p}R_{v}T^{2}_{500}}} 
\end{equation}

Here we focus on midlatitude extreme moist heat conditions where CAPE is in general high ($>$1000 J kg$^{-1}$). For those cases, the LFC is near the lower free troposphere and EL is close to the tropopause, and we approximate them by $p_{LFC}\simeq 950$ hPa and $p_{EL}\simeq 100$ hPa. The mean $T_{500}\simeq 260$ K gives $q^{*}_{500}\simeq 0.0028$ kg kg$^{-1}$. Substituting these typical values into Eq.\ref{eq7}, we provide a scaling for midlatitude high CAPE:

\begin{equation} 
\label{eq8}
CAPE\simeq 0.22(MSE_{s}-MSE^{*}_{500})
\end{equation}

Previous studies have explored the relationship between CAPE and environmental MSE, either under the radiative–convective equilibrium framework \cite{emanuel_bister_1996,romps_2016,Agard_Emanuel_2017} or through linear regression methods \cite{wang_moyer_2023}. In contrast, our derivation begins with the fundamental definitions of parcel buoyancy and CAPE, similar to \cite{Li_Chavas_2021} but incorporating minimal assumptions about the geometric characteristics of CAPE, to establish a robust linear relationship between CAPE and key atmospheric parameters near the surface and in the free troposphere. The coefficient in Eq.\ref{eq8} may vary with $p_{LFC}$, $p_{EL}$, and $T_{500}$, which can be precisely adjusted using the full scaling equation (Eq.\ref{eq7}) when applied to different regions or datasets. Here, the derived coefficient of 0.22 for high-CAPE environments over midlatitude land aligns closely with the linear regression results (0.22–0.29) reported for high CAPE over continental North America \cite{wang_moyer_2023}.

\subsection*{Data Availability}
The 3-hourly ERA5 reanalysis data \cite{NCAR_RDA_ERA5,Hersbach_etal_2020} from 1980--2022 are publicly available at \url{https://rda.ucar.edu/datasets/ds633.0/}. 

\subsection*{Code Availability}
The xcape python package is available at \url{https://github.com/xgcm/xcape/tree/master}. The Metpy python package is available at \url{https://unidata.github.io/MetPy/latest/index.html}.

}

\showmatmethods{} 

\acknow{This research is part of the MIT Climate Grand Challenge on Weather and Climate Extremes. Support was provided by Schmidt Sciences, LLC. We thank Kerry Emanuel, Paul O'Gorman, P. J. Tuckman, and Divya Rea for helpful discussions.}

\showacknow{} 

\bibliography{all-bibtex}

\end{document}